\documentclass{vgtc}  

\onlineid{0}

\vgtccategory{Research}

\vgtcpapertype{application/design study}

\title{See or Hear? Exploring the Effect of Visual/Audio Hints and Gaze-assisted Instant Post-task Feedback for Visual Search Tasks in AR}

\author {Yuchong Zhang \thanks{yuchong@chalmers.se; yuchongz@kth.se}\\
        \scriptsize Chalmers University of Technology, Sweden
\and Adam Nowak \thanks{203151@edu.p.lodz.pl}\\
     \scriptsize Lodz University of Technology, Poland
\and Yueming Xuan \thanks{echoxuan98@gmail.com}\\
     \scriptsize Chalmers University of Technology, Sweden
\and Andrzej Romanowski \thanks{androm@kis.p.lodz.pl}\\
     \scriptsize Lodz University of Technology, Poland
\and Morten Fjeld \thanks{fjeld@chalmers.se; Morten.Fjeld@uib.no}\\
     \scriptsize Chalmers University of Technology, Sweden \\ \scriptsize University of Bergen, Norway
}

\abstract{%
  Augmented reality (AR) is emerging in visual search tasks for increasingly immersive interactions with virtual objects. We propose an AR approach providing visual and audio hints along with gaze-assisted instant post-task feedback for search tasks based on mobile head-mounted display (HMD). The target case was a book-searching task, in which we aimed to explore the effect of the hints together with the task feedback with two hypotheses. \textbf{H1}: Since visual and audio hints can positively affect AR search tasks, the combination outperforms the individuals. \textbf{H2}: The gaze-assisted instant post-task feedback can positively affect AR search tasks. The proof-of-concept was demonstrated by an AR app in HMD and a comprehensive user study (n=96) consisting of two sub-studies, Study I (n=48) without task feedback and Study II (n=48) with task feedback. Following quantitative and qualitative analysis, our results partially verified \textbf{H1} and completely verified \textbf{H2}, enabling us to conclude that the synthesis of visual and audio hints conditionally improves the AR visual search task efficiency when coupled with task feedback.
}

\keywords{Augmented reality, Search task, Visual hint, Audio hint, Gaze assistance, Instant post-task feedback, User study}

\teaser{
 \centering
  \includegraphics[width=.95\linewidth]{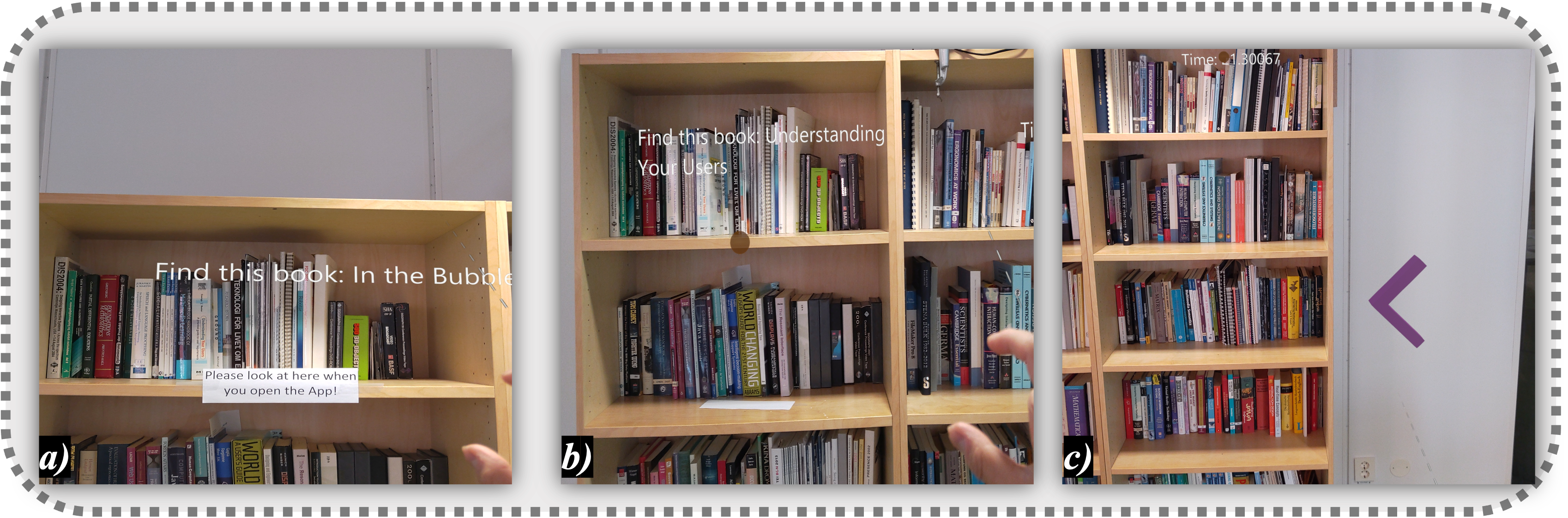}
  \caption{\scriptsize{Case study: Visual book-searching task with the aid of AR using HMD. $a)$: Our AR app gives a book title as a search stimulus. $b)$: Instant post-task feedback is provided by a dot smoothly following the eye trajectory. (Note: the dot is bright and highly visible in AR but is dim and difficult to see in print; here, it is located between the first and second shelves of the bookcase). $c)$: The visual hint, as a bright purple arrow, supports the task. The timer is designed for measuring the task time (at the top of this figure). Both the gaze playback and visual hints are rendered in real world space for precise displacement. Please check our supplementary video for the demonstration of the book-searching task.}}
  \label{fig:teaser}
}

\graphicspath{{figs/}{figures/}{pictures/}{images/}{./}} 

\usepackage{tabu}                      
\usepackage{booktabs}                  
\usepackage{lipsum}                    
\usepackage{mwe}                       

\usepackage{mathptmx}                  

\begin{document}


\firstsection{Introduction}
\label{intro}

\maketitle

Even as Augmented Reality (AR) continues to mature technically, it has already become a cutting-edge technique for ordinary use in everyday life \cite{pfeuffer2021artention,zhang2021supporting,zhang2021novel}. AR features the capacity to superimpose virtual context-related information onto the physical world in the form of graphics \cite{piening2021looking,qian2019portal,zhang2023playing,zhang2022site}. This technology is endorsed and applied in numerous fields due to its capability of providing real-time interaction, as well as generating interactive interfaces of visualized digital content \cite{lee2011design,nowak2021augmented,zhang2023virtuality}. An emerging trend of AR is to deploy this technique to generate capability for conducting specific tasks. The visual search task in AR is a widespread research area which has received considerable attention, while the representations of it can be diverse, including visual element searching \cite{huang2022effects}, real world object searching \cite{contreras2017semantic}, or general searching \cite{shaikh2021augmented,reardon2018come}. As a means to make the search process more accessible, it is even possible that HMD devices with the most advanced AR technology could soon become as ubiquitous as smartphones \cite{rivu2020stare} to empower the search process to become more accessible. We investigated the specialized case study of a book-searching task in AR, which includes all the essential components of a visual search procedure: visual environment, a particular object (the target book), and distractors (irrelevant books) \cite{treisman1980feature}, and is easy to realize.

When performing particular tasks in the context of AR, hints can improve the ultimate task performance. The type of hint, acting as a central and core tool, can vary. A widely-used tool is the visual hint \cite{volmer2018comparison}, which is a mixture of graphical representations in AR with user interface (UI) actions associated with the physical world \cite{white2007visual}. This effectively provides spatial and temporal guidance in AR. Another well-adopted hint is the audio hint, which presented as voice commands or directives can swiftly and clearly aid the user. A number of researchers have exploited audio hints for the purpose of informing, guiding, and directing within their designs \cite{lyons2000guided,sawhney2000nomadic}. However, even though there is a substantial body of research into how visual and audio hints work in AR tasks, little has focused on search tasks other than exploring the combination of visual and audio hints in one AR framework. In our study, we observe both the separate effect of these two types of hints and exclusively their combined effects in AR searching. The context of a visual book-searching task is engaged in an AR approach with supportive visual or/and audio hints.

Gaze assistance, in particular, has a promising role in AR applications owing to its easy, natural, and fast way of involving people in virtual environments \cite{park2008wearable}. Gaze-assisted techniques and implementations have been demonstrated to efficiently support AR applications by providing better user experiences (UX) \cite{lankes2016gazear} and more accurate target selection techniques \cite{kyto2018pinpointing}. Human eyes are easy to track, and gaze implicitly conveys what people are interested in. Thus, eye-tracking approaches are becoming more pervasive in interactive AR devices, which are mainly mobile head-mounted or hand-held displays (HMDs and HHDs) \cite{zhang2023need}, such as Microsoft HoloLens or Magic Leap \cite{piening2021looking}. Human gaze is utilized for adjusting and adapting virtual information that is being projected onto the real world. The usage of gaze has been extensively demonstrated to be efficient in visual search tasks. Noteworthily, eye gaze recordings have been harnessed in many AR-led or AR-guided contexts \cite{wolf2021gaze,seeliger2021exploring,oney2020evaluation}. However, very little research has investigated the combination of visual/audio hints and human gaze within visual search tasks within the same context.

Another imperative aspect of AR is task feedback. At present, there is almost no in-depth research into involving gaze assistance for task feedback. It is crucial for domain users to obtain better task performance in many situations related to AR \cite{sousa2016sleevear,clemente2016humans}. According to Vieira et al. \cite{vieira2015augmented}, task feedback has the potential to provide more specific information needed by the users in AR, which encouraged us to explore its potential influence in the same context. Here, we explore the effect of task feedback through gaze-assistance given that both of these have proven their efficacy in AR while there is an evident gap on the combination of them. Instead of real-time feedback \cite{petersen2013real}, we adopted post feedback \cite{zhao2020intelligent,scolere2022building} in our visual book searching context since it allows people to see their task reflection after a complete searching process \cite{anderson2013youmove}. The task feedback in our study is embodied as the playback of the human eye gaze recorded during the book-searching process, which is symbolized as an instant evaluation method. A vision-friendly colored dot smoothly following the path is used to display the eye trajectory (Figure ~\ref{fig:teaser}.$b$). Specifically, the gaze playback component enabled by eye-tracking serves as the instant post-task feedback in our study design. It is noteworthy that no relevant research, to our knowledge, has studied the effect of associating visual/audio hints with instant post-task feedback in the same context of AR visual searching.

By realizing these research gaps (combining visual and audio hints; task feedback and gaze assistance; visual/audio hints and task feedback), this paper proposes a complete AR approach for visual search tasks, examines the independent and the combined impact of visual and audio hints, and studies instant post-task feedback through gaze assistance. Under the premises that visual/audio hints yield beneficial effects \cite{arevalo2021assisting,cidota2016comparing,marquardt2019non}, so as task feedback \cite{zahiri2018evaluation,clemente2016humans}, through the AR book-searching task, we intend to test the hypotheses: 

\begin{itemize}
    \item \textbf{H1}: Since visual and audio hints have a positive effect in facilitating AR visual searching performance and decreasing perceived workload in AR, the combination of these two hints has a greater effect than either does individually.
    \item \textbf{H2}: The gaze-assisted instant post-task feedback has a positive effect for task performance and perceived workload reduction in AR visual search tasks. 
\end{itemize} 

To examine \textbf{H1} and \textbf{H2}, we designed a comprehensive between-subject \cite{yip2019improving} user study with a within-subject factor \cite{charness2012experimental} followed by statistical analyses with the aid of a specialized-for-searching bookcase. The paper contributes:

\begin{itemize}
    \item Proposing an AR approach supporting visual and audio hints, as well as gaze-assisted instant post-task feedback for visual search tasks based on mobile HMD;
    \item Exploring the effect of combining visual and audio hints in contextual AR visual searching;
    \item Exploring the effect of instant post-task feedback through gaze assistance in the same context.
    \item Exploring the effect of combining hints and instant post-task feedback in a same AR context.
\end{itemize}

This paper is organized as follows. Section~\ref{rw} describes related work of AR with gaze assistance, visual/audio hints, and task feedback. In Section \ref{sd}, we introduce the details of our proposed AR approach. We also incorporate the particular use case of the visual book-searching task. The entire user study is presented in Section~\ref{us}, with two sub-studies conducted. Section~\ref{re} contains the results from the study and quantitative data analysis along with the qualitative summary. Discussion with limitations follows in Section~\ref{dis}, and conclusions and future work are elaborated in Section~\ref{con}.


\section{Related Work}
\label{rw}

\subsection{Visual Search Tasks in AR}
Visual search tasks are becoming more common in AR because of the capability of projecting additional virtual information on the physical environment. Contreras et al. \cite{contreras2017semantic} presented a mobile application with AR encapsulated to admit users to search for desired places, people or events on a university campus. The superiority of AR lay in the fact that it offered certain visual elements that helped users to better locate the required result. Rafiq and colleagues \cite{rafiq2014secure} proposed a dynamic AR framework to support an online book-searching task by using mobile augmented data. This framework also introduced a security layer which ensured the protection of sensitive cloud data. Gebhardt et al. \cite{gebhardt2019learning} utilized gaze movement data to observe the MR object's label in a visual searching process through a reinforcement learning method. Trepkowski et al. \cite{trepkowski2019effect} presented a series of simulating experiments to investigate how visual search performance is affected by the field of view and information density in AR, indicating that a significant effect was caused by these two factors. Van Dam et al. \cite{van2020drone} studied the cues in a drone-based AR signal detection task but found no significant differences across AR cue types. Nevertheless, the effects of hints and task feedback continues to be more valued in AR visual searching.

\subsection{Hints in AR}
Numerous researchers have endeavored to bring different modalities of hints into AR/XR systems as auxiliary tools for reaching desired results. Of these, visual hints \cite{white2007visual,nonino2021subtle,behringer2002model,zhu2014ar,zhao2016cuesee,zhang2022initial} and audio hints\cite{lyons2000guided,zhu2014ar,lindeman2007hear,sundareswaran20033d} are the two most common representations used. Arboleda et al. \cite{arevalo2021assisting} presented augmented visual hints in a robot-involved AR system which aimed to enhance the visual space of the robot operator about the position of the robot gripper in the workspace, where the visual hints were used to improve distance perception and then the manipulation and grasping task performance. For tangible AR, White et al. \cite{white2007visual} examined visual hints to enable discovery, learning, and completion of gestural interaction in a tangible environment. Seven visual hint types were generated: text, diagram, ghost, animation, ghost+animation, ghost+text, and ghost+text+animation. Two decades ago, Sawhney et al. \cite{sawhney2000nomadic} harnessed audio information to keep users of wearable devices updated with incoming messages and events. Lyons et al. \cite{lyons2000guided} developed an AR game system named "Guided by Voices", which equipped the user with a narrative sound clip that indicated the scenarios encountered and the correct steps to be taken for proceeding. Interestingly and more recently, Mulloni et al. \cite{mulloni2011user} found that AR can be integrated with a mobile navigation system, while audio hints can be advisable for eye-free usage. Cidota et al. \cite{cidota2016comparing} compared the effects of automatic visual and audio notifications regarding workspace awareness in AR remote collaborating. However, the integration of visual and audio hints in one AR context is still insufficient.

\begin{figure}[h]
  \centering
  \includegraphics[width=.85\linewidth]{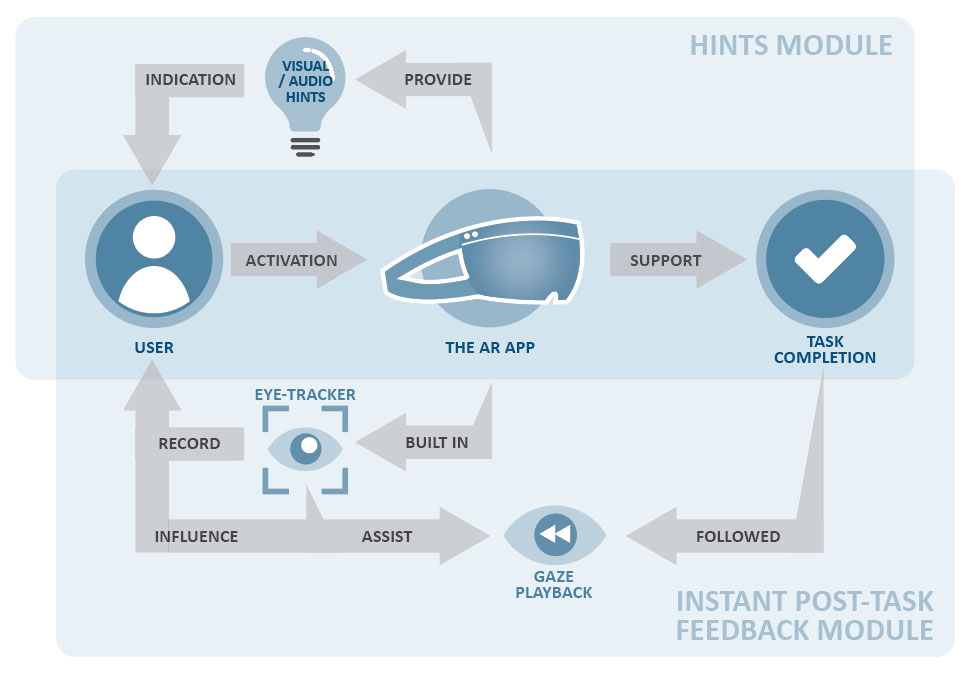}
  \caption{\scriptsize{Block diagram: The proposed AR approach with the two core inter-correlated modules: hints (top) and instant post-task feedback (bottom). The AR app builds on the intersection).}}
  \label{fig:bd}
\end{figure}

\subsection{AR with Gaze Assistance}
Some research has already pioneered gaze-assisted UI to access more contextual information \cite{ajanki2011augmented,lu2020glanceable,pathmanathan2020eye,sasikumar2019wearable}. It has been shown that the optical see-through (OST) HMD which harnesses human gaze with eye-tracking as the interaction metaphor can contribute to efficient results \cite{looser2007evaluation}. Interestingly, there are some studies exploring eye-tracking to present menus to aircraft pilots, and adjust their contents based on what the pilot is looking at \cite{safety4010008}. In 2005, Curatu et al. \cite{curatu2005projection} proposed a novel conceptualized system adding eye-tracking capabilities to a Head-Mounted Projection Display (HMPD), which was satisfyingly performed from a low-level optical configuration. Three years later, Park et al. \cite{park2008wearable} pioneered a system which includes a Wearable Augmented Reality System (WARS) to examine their proposition on an experiment in which they selected desirable items in an AR gallery with content mobility. Rivu et al. \cite{rivu2020stare} successfully demonstrated the superiority of eye-tracking, showing that users are more positive in concentrating on their ongoing conversations when there is more gaze interactivity under AR environment. All the relevant research proved that the gaze assistance in AR activates improved perception of people.

\subsection{Task Feedback in AR}
Task feedback evaluation in AR has received compelling attention in the past years. A recent study carried out by Liu et al. \cite{liu2012evaluating} evaluated the influence of real-time task feedback in mobile handheld AR, which suggested that significant benefits will emerge if task feedback is engaged during the AR task. Zahiri et al. \cite{zahiri2018evaluation} studied the feedback in AR for supporting a surgical training, finding that most of the users preferred receiving the feedback while the task was being performed. Murakami et al. \cite{murakami2013poster} found that haptic feedback as task evaluation can help users perform more effectively in a wearable AR system for virtual assembly tasks. Clemente et al. \cite{clemente2016humans} demonstrated that continuous visual AR feedback can deliver effective information for the users in their sensorimotor control with a robotic hand, which has implications for amputee assistance in a clinical scenario. Anderson et. al designed a system called YouMove which used itnteractive At mirrors to train users to record and
learn physical movement sequences, where they employed post-stage feedback in their user study. Nonetheless, research regarding post feedback in AR is still scarce. Even though some researchers have identified the effects of instant post-task feedback in gaze-interaction embodied virtual reality (VR) \cite{riegler2020gaze}, the gap of linking it with gaze assistance in AR is still obvious, as is the gap of combining it with visual/audio hints. Our study intended to fill the aforementioned gaps and open up future research directions regarding AR visual search tasks. 

\section{System Design}
\label{sd}

\subsection{Outline of the Proposed AR Approach}
The proposed system, featuring an AR app, is an interactive tool for users carrying out a book-searching task. The equipment kit consists of a mobile HMD with built-in eye tracker which is responsible for recording the user's gaze during the task being conducted. An overall block diagram of our proposed system is shown in Figure \ref{fig:bd}. There are two inter-correlated modules: the hints module and the instant post-task feedback module. The first module was designed to address \textbf{H1,} while the second targeted \textbf{H2}. The components "User", "The AR app", and "Task completion" are commonly owned by the two modules. The AR app is the core of the system which bears the necessary information needed for the book-searching task including hints and instant post-task feedback. The hints module features a "Visual/audio hints" component while "Eye-tracker" and "Gaze playback" components characterize another module. During the task period, the user's eye motion is entirely captured by the eye tracker and simultaneously recorded. These recordings then act as the feedback for later self-reflection. After the completion of an assigned task, users can obtain data on the total time elapsed and then review the gaze playback to monitor their performance and make improvements for the next task. The central advantages of this system are the simultaneous visual/audio hints designed to enhance task performance, and the instant post-task feedback pathway which offers users the opportunity to make immediate improvements. 

\subsection{Book-searching Task}
\begin{figure}[h]
\centering
  \includegraphics[width=.85\linewidth]{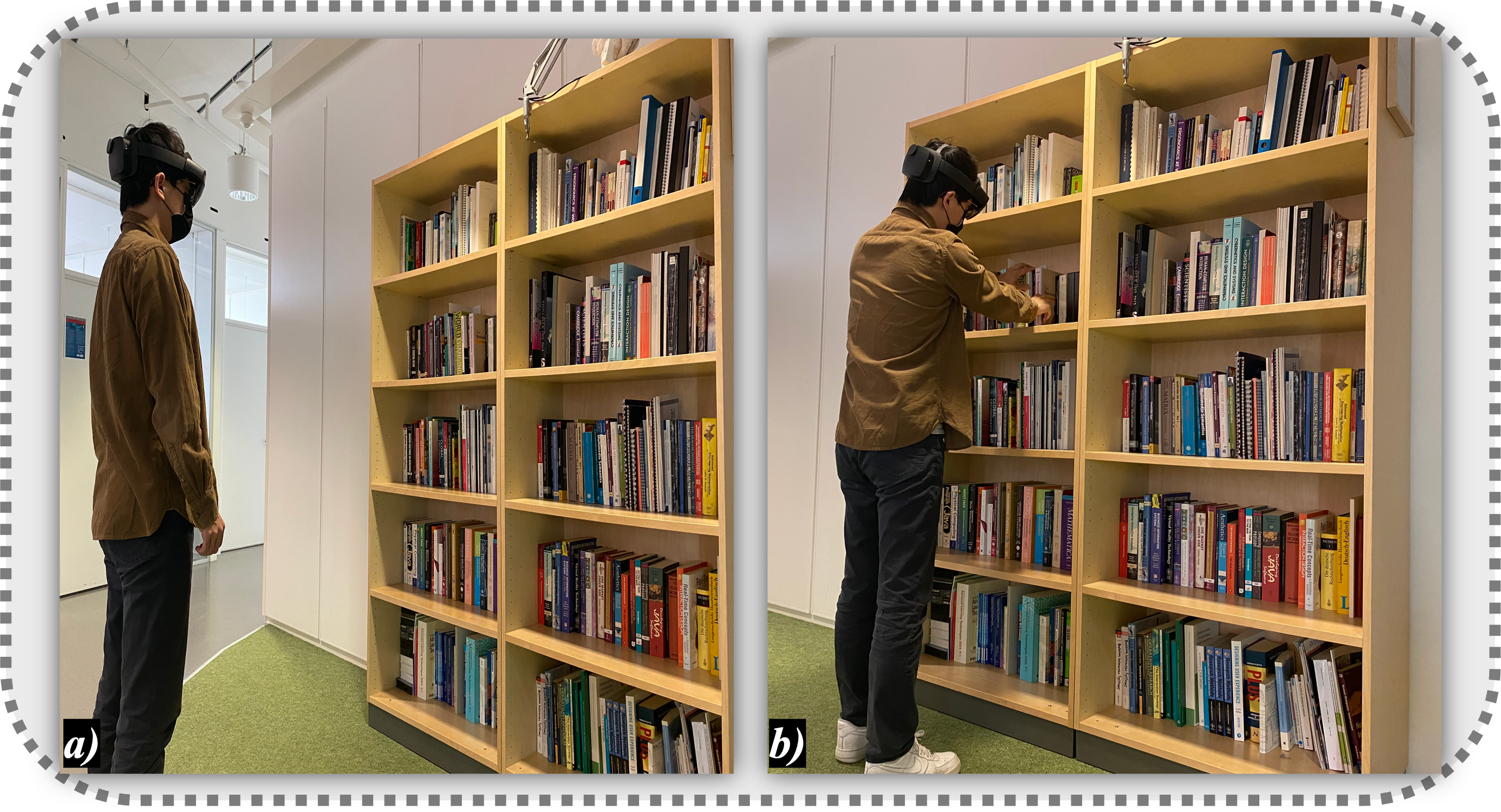}
  \caption{\scriptsize{User study: An example scene of one participant searching for a book. $a)$: During searching; $b)$: Book found. The two congruent bookcases are used for the two searching tasks.}}
  \label{fig:us_par}
\end{figure}

The book-searching task was implemented with the aid of the proposed AR approach. The AR app encoded a number of book titles which were used as the stimulus in the study. The user with an HMD with a real-time built in eye-tracker would stand in front of the bookcase, to find a specific but randomly determined book (Figure \ref{fig:teaser}.$a$) after activating the app. In the beginning, users allocated in distinct sub-studies were assigned into particular groups. Depending on the settings, different types of hints with/without instant post-task feedback were then provided. There were two identical book-searching tasks in this study and the gaze playback served as the feedback which was inserted between the two tasks. To precisely control the experimental variables while differentiating the task content, we adopted two congruent bookcases but with totally different books being cleared placed. (Figure \ref{fig:us_par}). The books used for the two tasks are also distinct. The spatial layout of books remained unchanged and unmoved in both bookcases throughout the entire study.

\subsection{Hints and Instant Post-task Feedback}
According to Arboleda et al. \cite{arevalo2021assisting}, visual hints can improve depth perception, which allows people to cognitively better understand their spatial environment. In Wolf et al.'s work \cite{wolf2018care}, animated arrows produced by an AR HMD were used as visual hints to assist dementia patients with navigation. We also chose arrows as the indicators, but showed them in an easily-recognizable color and in a static state. Since all virtual artifacts are displayed in the world space, users' movements do not affect an object's position. As shown in Figure \ref{fig:teaser}.$c$, the bright-purple arrow floating in the middle of a shelf visually helps the user to locate the target book. This visual arrow is designed to appear within the users' field of view, easily observed by users wearing the HMD. The arrow for the first searching task was positioned on the left side while it appeared on the right side, since the first task started from the left bookcase. Furthermore, the audio hint also plays a decisive role in guiding users during the searching process. Marquardt et al. \cite{marquardt2019non} demonstrated that AR users can be guided with higher accuracy and within a shorter searching time by incorporating auditory hints. In our developed app, users receive an audio instruction about the approximate location of the targeted book. The auditory information is presented as a clear and simple instructive voice message, allowing the users to finish the task with higher efficiency. This message was merely given once at the beginning of the task to serve as the means of guiding, and not repeated during the latter searching process to avoid being as the external noise disturbance. 

The functionality and advantages of task feedback have been described in Cao et al.'s findings \cite{cao2019ghostar}, that furnishing feedback in AR environments offers users more awareness of surrounding information, which ensures task efficiency and correctness. Appropriate feedback in a visually distinguishable form can alleviate their workload and result in positive improvements in task performance \cite{paelke2014augmented}. In our study, the instant post-task feedback was provided as a visible playback of gaze recording. As shown in Figure \ref{fig:teaser}.$b$, the user's gaze was denoted as a brightly colored dot following the trajectory. During the book-searching phase, gaze trajectories were marked and recorded. After one search task, users then watched the playback of their trajectories, after which they proceeded to the next task.

\subsection{Apparatus}
We used Microsoft HoloLens 2 glasses and the dedicated AR app we developed. The mobile headset weighs 566 grams and has a built-in battery. The see-through holographic lenses, SoC Qualcomm Snapdragon 850, are a second-generation custom-built holographic processing unit with 4 GB RAM memory and 64 GB storage. The users use the glasses freely without any connection to a power supply or external computing device. Built in eye-trackers allow us precise recording of users' gaze trajectories. The app was developed with use of Unity3D 2020.3.20f1 and Mixed Reality Toolkit (MRTK).

\section{User Study}
\label{us}
We designed and implemented a comprehensive comparative user study based on the between-subject with coupling within-subject factors. The independent variables were the tasks (within-subject) and the groups (between-subject). The study was comprised of two sub-studies (Study I and Study II) where the participants from Study II received the instant post-task feedback while those from Study I did not. Both studies I and II followed the same between-/within-subject design. Before starting the experiment, we conducted a pre-testing session where several people were invited to test the desired functionalities of the AR app. Some minor adjustments were then made to improve the visual and locational clarity, including the visual hint made into a more vision-friendly arrow, and our testing environment moved to a more spacious and bright function room.

\begin{figure}[ht]
\centering
  \includegraphics[width=\linewidth]{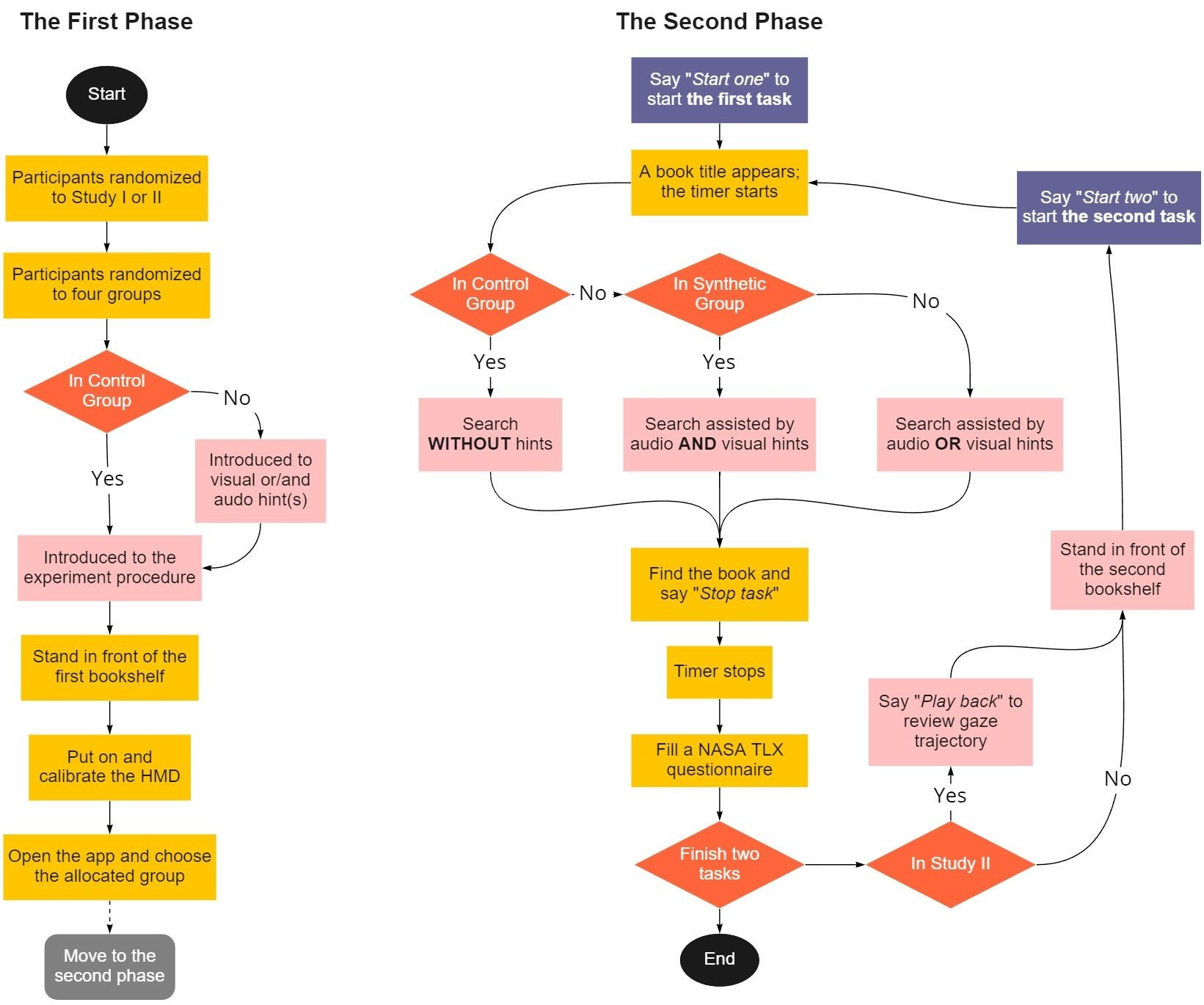}
  \caption{\scriptsize{User study: The flowchart of the task procedure. The two phases show how the participants were engaged in the study. The first phase presents the prerequisite of the study while the second phase illustrates the formal searching tasks.}}
  \label{fig:us_flow}
\end{figure}

\subsection{Participants}
We recruited 96 participants (different from the people involved in the pre-testing session), of whom most were from a local university. Their ages ranged from 19 to 57 ($mean$ = 27.92, $SD$ = 7.70); 54 voluntarily self-reported male and 42 self-reported female. There were 48 participants in both Study I (24 males and 24 females, aged from 20 to 44 ($mean$ = 26.28, $SD$ = 4.64) and Study II (30 males and 18 females, aged from 19 to 57 ($mean$ = 29.11, $SD$ = 9.20) respectively. No participants decided to quit during the experiment, reported any forms of color blindness, or showed discomfort or rejection of the HMD used in this study. Each participant carried out the experiment independently and correctly. It is therefore reasonable to base our analysis and arguments on the data obtained. The whole study (including I and II) was executed in a bright and spacious function room without any external distractions other than the two bookcases used. There were no additional noises affecting the participants during the whole experimenting process. The acoustic conditions of the function were satisfactory to ensure everyone clearly receiving the audio hint. Each participant was given a small gift for helping with the study afterwards. All of the data collection corresponded to the Covid-19 and ethical rules of the authors’ home universities.


\subsection{Experiment Design}
The participants (n=96) were randomly sorted into Study I and Study II (n=48 in each) and again randomly and evenly distributed in four groups in both studies: control, audio, visual, and the combined groups (n=12 in each). The only variation among the groups was the hint difference. We utilized identical settings in the two sub-studies; however, all the groups from Study II received the instant post-task feedback while those from Study I received no such feedback. All participants (n=96) executed two book-searching tasks. For the control group, we did not offer any hints. The visual and audio groups received visual (the directive bright-purple arrow pointing out a specific bookcase shelf) or audio (the instructive and articulate voice message (\textit{"Please look at the Xth shelf of the bookcase from the top"})) hints respectively. Finally, the combined group was given both sets of hints when performing the book-searching. The AR app was developed by incorporating the precise calculation of the inherent parameters of the bookcase (for example, the height of each shelf). The height of the participants did not affect the results since the visual arrows were continuously pointing out the correct shelf. The group allocation was employed for controlling the variables to verify \textbf{H1} and the hints module in the block diagram (Figure \ref{fig:bd}). Each group in Study II underwent the feedback phase, where all participants were requested to watch their own gaze recordings from their first book-searching task before undertaking the second task. This was used for identifying the exact role of the instant post-task feedback. The results were to address \textbf{H2}, as well as the usefulness of the instant post-task feedback module \ref{fig:bd}. Everyone was requested to search for different books in two different bookcases in the respective two tasks (Figure \ref{fig:us_par}).

\subsection{Task Procedure}
The duration time for the study of every participant lasted from 10 to 15 minutes. A short and concise pre-training session of book-searching was employed to get all the participants familiarised with the entire environment to maximally eliminate the potential accumulative learning effect among the tasks. The task completion time (TCT), which is defined as the duration from the participant initiates the searching until the book has been found, was measured regarding each task. The complete procedure for our study (including studies I and II), as seen in Figure \ref{fig:us_flow}, consisted of two phases:

\begin{itemize}
    \item 
    \small{The \textbf{first phase} started with the introduction of the study and calibration of the HMD. During the introduction, each participant was randomly allocated to one of the four groups followed by an explanatory overview of the experiment procedure by conductor of the study. If participants were not assigned to the control group, the roles of the audio or/and visual hint(s) were introduced and explained. Participants were then told to stand in front of the first test bookcase at a distance of one metre (a marker was stuck on the ground). Then they were instructed to put on and calibrate the HMD. Next, participants opened the designated AR app, and underwent the pre-training of a complete book-searching procedure without receiving any hints/feedback. They then used the floating virtual buttons for choosing their specified groups with a voice prompt: "xx groups is selected".}
    \item \small{In the \textbf{second phase}, the participants started the first book-searching task by saying "Start one". As the title of a randomly selected book appeared (Figure \ref{fig:teaser}.$a$) and a built-in timer began (on the top of Figure \ref{fig:teaser}.$c$), they started searching based on the given title. In this phase, all participants from studies I and II except those from the control groups received either visual/audio hint(s) or the combined set, which pointed out a specified bookcase shelf. The visual hint remained in the real world space regardless of the participants' position. The audio hint was, however, not repeated. Upon successfully finding the book, the participants completed the first search task by saying "Stop task" to stop the timer while the TCT of this task was measured and noted. Next, only the participants from Study II began the review of their gaze playback by articulating "Playback". On finishing, they filled out a NASA Task Load Index (TLX) questionnaire \cite{hart1988development}, a widely-used assessment tool for perceived workload. The participants were then directed to stand one metre in front of the second test bookcase with saying "Start two" and "Stop task" to start and complete the second task using the same process. The sign of the book found in each task is when the user successfully visually locate the book by seeing its title, and then stop the timer orally. All of the voice commands given by participants to the device were recognized at first try during the formal study because of the quiet environment and they knew how to stop the timer by speech from the pre-training sessions. Finally, they were given a second NASA TLX questionnaire to fill out. All participants were told to unrestrictedly express their feedback after completing the study.} 
\end{itemize}

\section{Results and Analyses}
\label{re}
Below we present the results from studies I and II. All participants (n=96) completed the study successfully in finding the correct books at first try in each task with different durations. Thus, no tasking accuracy but the tasking time was harnessed. The dependent variables employed for assessing the tasks were TCT and NASA TLX. 

\begin{figure}[!ht]
\centering
  \includegraphics[width=\linewidth]{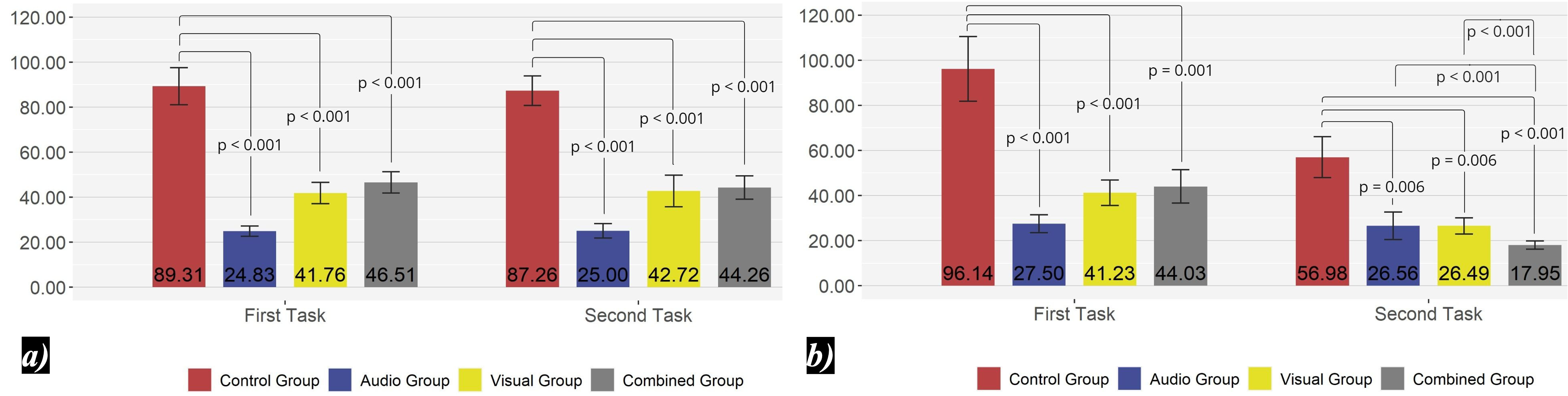}
  \caption{\scriptsize{Study I ($a$: \textit{without instant post-task feedback}) and Study II ($b$: \textit{with instant post-task feedback}): The mean TCT ($s$) used per task by the four groups. Error bars show mean $\pm$ standard error (SE).}}
  \label{fig:I_II_time}
\end{figure}

\subsection{Study I: Without Instant Post-task Feedback}

\subsubsection{Task Completion Time}
We first present the results according to the measured TCT of the two book-searching tasks conducted on task performance. As shown in Figure ~\ref{fig:I_II_time}.$a$, the descriptive statistics show that in the first task the mean TCT for the control group ($mean$ = 89.31; $SD$ = 29.80) was much longer than for the other three experimental groups: audio group ($mean$ = 24.83; $SD$ = 8.37), visual group ($mean$ = 41.76; $SD$ = 17.10), combined group ($mean$ = 46.51; $SD$ = 17.04). Similarly in the second task, the mean TCT of the control group ($mean$ = 87.26; $SD$ = 23.61) was still considerably higher than the other groups. Notably, the audio group ($mean$ = 25.00; $SD$ = 11.70) still had the shortest TCT in comparison with the visual group ($mean$ = 42.72; $SD$ = 25.36) and the combined group ($mean$ = 44.26; $SD$ = 18.71). Also notably, the combined group did not have an obvious TCT decrease compared to the other three groups in Study I.

To identify the significant differences of the results, a 2 (task) * 4 (group) mixed analysis of variance (ANOVA) ($p$ = 0.05) was performed given that the normality of the data was affirmed. This analysis was performed using IBM SPSS Statistics, as were the following statistical measurements. Bonferroni-corrected post hoc tests were employed to determine if the pairwise groups were significantly different. We found that all participants did not have statistically significant shorter TCT ($mean$ = 49.81; $SD$ = 30.51) in the second task compared to the first ($mean$ = 50.60; $SD$ = 30.61), $F$(1, 44) = 0.183, $p$ = 0.671.  But there were significant main effects of the four groups on the TCT measured ($F$(3, 44) = 21.042, $p$ $<$ 0.001). No significant interaction was found between the tasks and the groups ($F$(3, 44) = 0.137, $p$ = 0.937). The Bonferroni post hoc tests showed statistical significance between every pairwise comparison of the control group with all the other groups in both tasks. For interpretation, the control group had significantly the most TCTs compared to the audio group ($p$ $<$ 0.001), the visual group ($p$ $<$ 0.001), and the combined group ($p$ $<$ 0.001). No significance was found among the pairs between the audio, visual, and combined groups.

\subsubsection{NASA TLX}
\begin{figure}[ht!]
\centering
  \includegraphics[width=\linewidth]{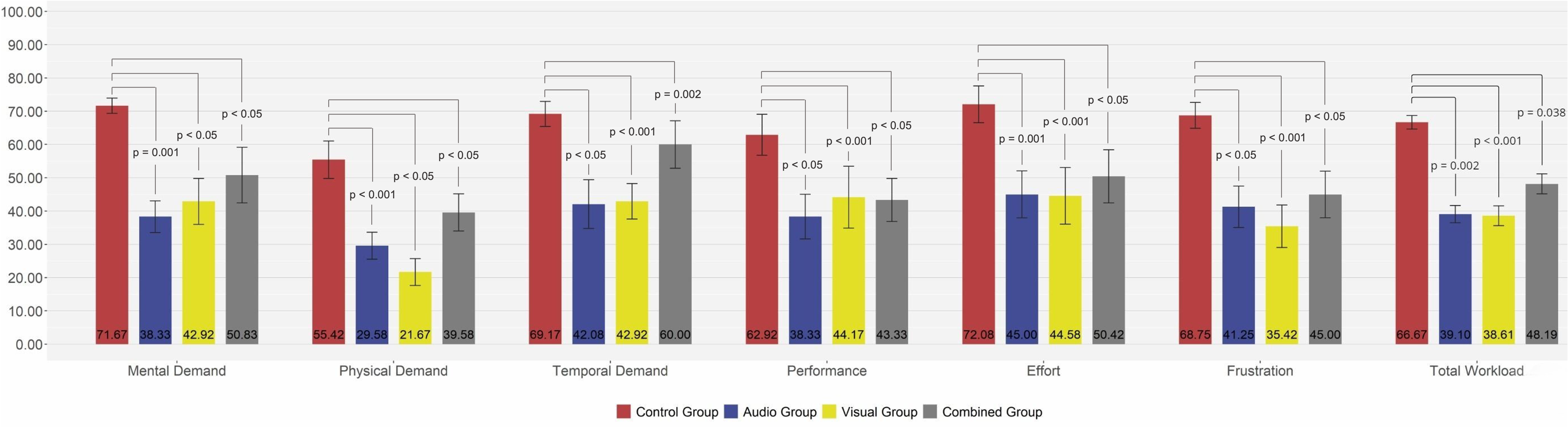}
  \caption{\scriptsize{First task in Study I: NASA TLX results of control (left) and three experimental groups. Error bars show mean $\pm$ standard error.}}
  \label{fig:I_nasa_task_1}
\end{figure}

\begin{figure}[ht!]
\centering
  \includegraphics[width=\linewidth]{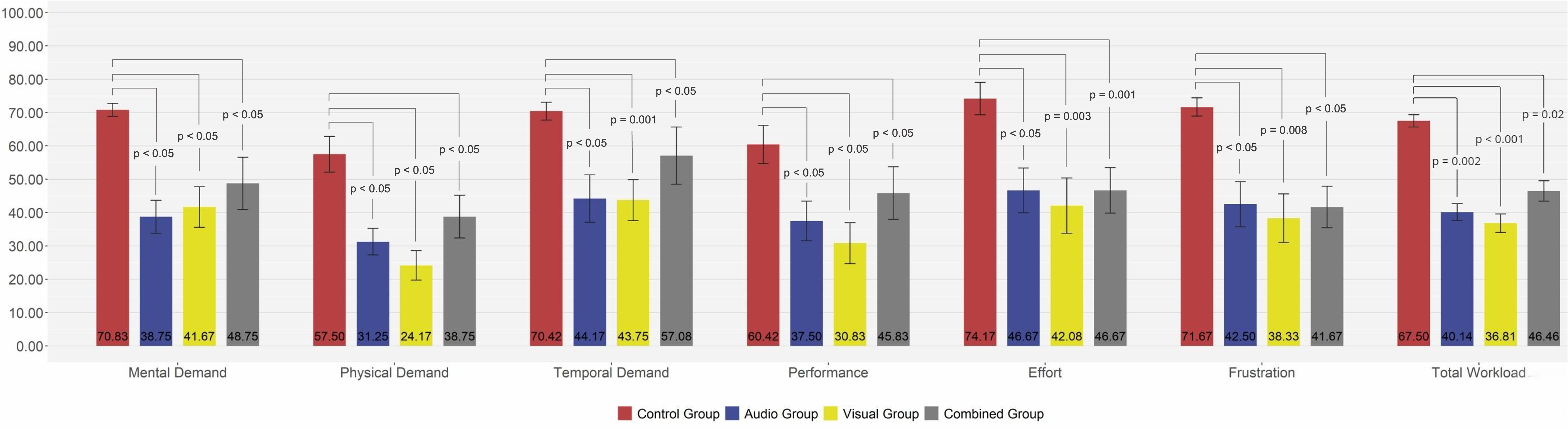}
  \caption{\scriptsize{Second task in Study I: NASA TLX results of control (left) and three experimental groups. Error bars show mean $\pm$ standard error.}}
  \label{fig:I_nasa_task_2}
\end{figure}

Here, we report NASA TLX answers gathered from the 48 participants from Study I on perceived workload. The descriptive statistics and visualizations of the collective data are shown in Figures ~\ref{fig:I_nasa_task_1} and ~\ref{fig:I_nasa_task_2}, where the mean values with standard errors (SE) are reported. By observing the Total Workload (average value calculated from the six indexes included: Mental Demand, Physical Demand, Temporal Demand, Performance, Effort, and Frustration), we found that the values of this metric from the control group (Total Workload: $mean$ = 66.67; $SD$ = 17.42 in the first task; $mean$ = 67.5; $SD$ = 15.66 in the second task) were greater than the other three groups for both tasks. Furthermore, participants in the control group yielded the most perceived workload in every index upon every task; in particular, the members of the control group had a substantial lead in Mental Demand ($mean$ = 72.67; $SD$ = 7.99 for the first task; $mean$ = 70.83; $SD$ = 6.72 for the second task) and Effort ($mean$ = 72.08; $SD$ = 19.20 for the first task, $mean$ = 74.17; $SD$ = 16.81 for the second task).

Since the NASA TLX values showed variance homogeneity, the 2*4 mixed ANOVA ($p$ = 0.05) was performed to examine the TLX results. The study showed that all participants did not have a significantly different perceived workload ($mean$ = 47.73; $SD$ = 21.39) in the second task compared to the first ($mean$ = 48.14; $SD$ = 20.71), $F$(1, 44) = 0.052, $p$ = 0.821. In contrast, there were significant main effects of the NASA TLX results among the four groups ($F$(3, 44) = 7.591, $p$ $<$ 0.001). No significant difference was found on the interaction between the tasks and the groups ($F$(3, 44) = 0.182, $p$ = 0.908). The post hoc test showed all three groups were significantly different when compared to the control group in the both tasks. We conducted statistical analysis on the six indexes and Total Workload. However, we only report the results of Total Workload in the following parts of this paper since the significance found on the original six indexes can be referred to in Figures \ref{fig:I_nasa_task_1} to \ref{fig:II_nasa_task_2}. The participants from the control group felt that there was significantly more perceived workload than those from the audio group ($mean$ = 39.09; $SD$ = 21.82; $p$ = 0.002), the visual group ($mean$ = 38.61; $SD$ = 25.48; $p$ $<$ 0.001), and the combined group ($mean$ = 48.19; $SD$ = 25.58; $p$ = 0.038). Similarly in the second task, the control group ($mean$ = 67.50; $SD$ = 9.25) had statistical significance between the audio ($mean$ = 40.14; $SD$ = 19.17; $p$ = 0.002), the visual ($mean$ = 36.81; $SD$ = 19.81; $p$ $<$ 0.001), and the combined groups ($mean$ = 46.46; $SD$ = 22.08; $p$ = 0.02). Again, no significant differences between every pairwise comparison of audio, visual, and combined groups were found in both tasks in Study I.

\subsection{Study II: With Instant Post-task Feedback}

\subsubsection{Task Completion Time}
As with Study I, we first considered the results by TCT for the two book-searching tasks performed with the instant post-task feedback in between. As shown in Figure \ref{fig:I_II_time}.$b$, the descriptive statistics also show that in the first task the mean TCT for the control group ($mean$ = 96.14; $SD$ = 49.43) was much longer than for the other three experimental groups: audio group ($mean$ = 27.50; $SD$ = 13.87), visual group ($mean$ = 41.23; $SD$ = 19.65), combined group ($mean$ = 44.03; $SD$ = 25.89). Similarly in the second task, the mean TCT of the control group ($mean$ = 56.98; $SD$ = 31.59) was still considerably higher than the other groups. However, the combined group ($mean$ = 17.95; $SD$ = 6.36) had the shortest TCT in comparison with the audio group ($mean$ = 22.91; $SD$ = 20.89) and the visual group ($mean$ = 26.49; $SD$ = 12.51). This result differed from that of the first book-searching task.

Likewise, after the normality of the TCT data was determined, the 2*4 mixed ANOVA ($p$ = 0.05) revealed that all participants had significantly shorter TCTs ($mean$ = 31.99; $SD$ = 25.33) in the second task compared to the first ($mean$ = 52.23; $SD$ = 40.49), $F$(1, 44) = 13.421, $p$ $<$ 0.001. The results also showed significant main effects of the four groups on the measured TCT ($F$(3, 44) = 14.390, $p$ $<$ 0.001). In addition, there was a significant interaction between the tasks and the groups ($F$(3, 44) = 2.844, $p$ = 0.048). The Bonferroni post hoc tests showed significant differences between every pairwise comparison of the control group with all the other groups in the first task. For interpretation, the control group had the longest TCT compared to the audio group ($p$ $<$ 0.001), the visual group ($p$ $<$ 0.001), and the combined group ($p$ = 0.001). No significance was found among the pairs between the audio, visual, and combined groups. Likewise, the second task also showed significant differences between the control group and the other three groups ($p$ = 0.006 for the audio and visual groups, $p$ $<$ 0.001 for the combined group). However, we found that the pairs of the audio-combined ($p$ = 0.001) and the visual-combined ($p$ = 0.001) had significance as well. The TCT of the combined group was statistically different from the other three groups in the second task.

By observing the time outcomes from both tasks, we found that part of our \textbf{H1} -- individual equipment of visual and audio hints have a positive effect in facilitating visual search task performance in AR -- conformed to the obtained results. But another part of \textbf{H1} -- the combination of visual and audio hints has greater effect than either does individually in AR searching -- was only valid in the second task. Furthermore, the time used in the first task (without instant post-task feedback) exceeded that in the second task (with instant post-task feedback) in every group. Hence, our \textbf{H2} -- Instant post-task feedback contributes to better performance in AR search tasks -- was also reflected according to the TCT results.

\subsubsection{NASA TLX}
\begin{figure}[ht!]
\centering
  \includegraphics[width=\linewidth]{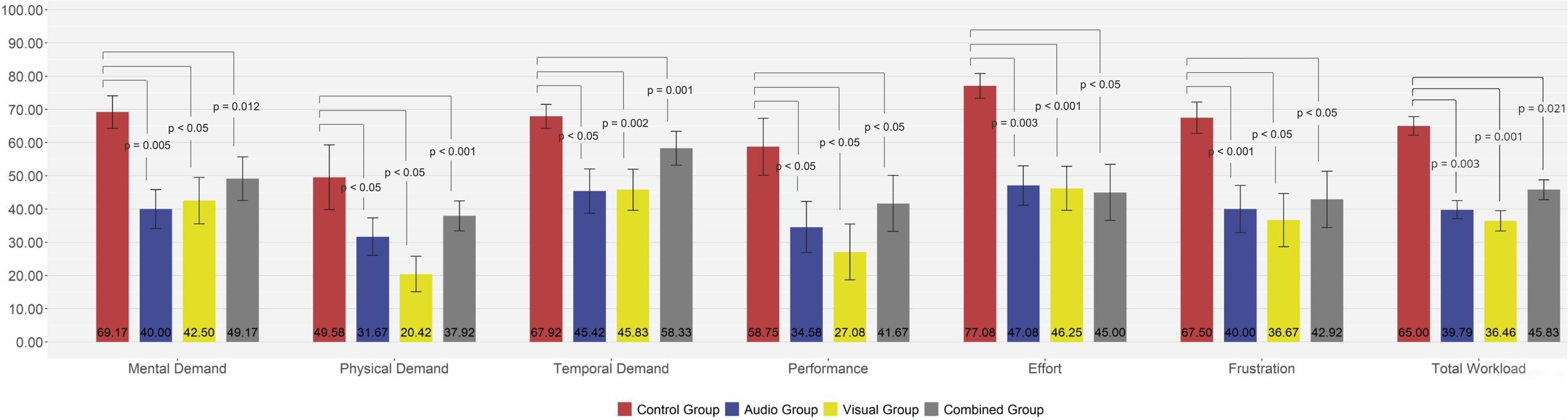}
  \caption{\scriptsize{First task in Study II: NASA TLX results of control (left) and three experimental groups. Error bars show mean $\pm$ standard error.}}
  \label{fig:II_nasa_task_1}
\end{figure}

\begin{figure}[ht!]
\centering
  \includegraphics[width=\linewidth]{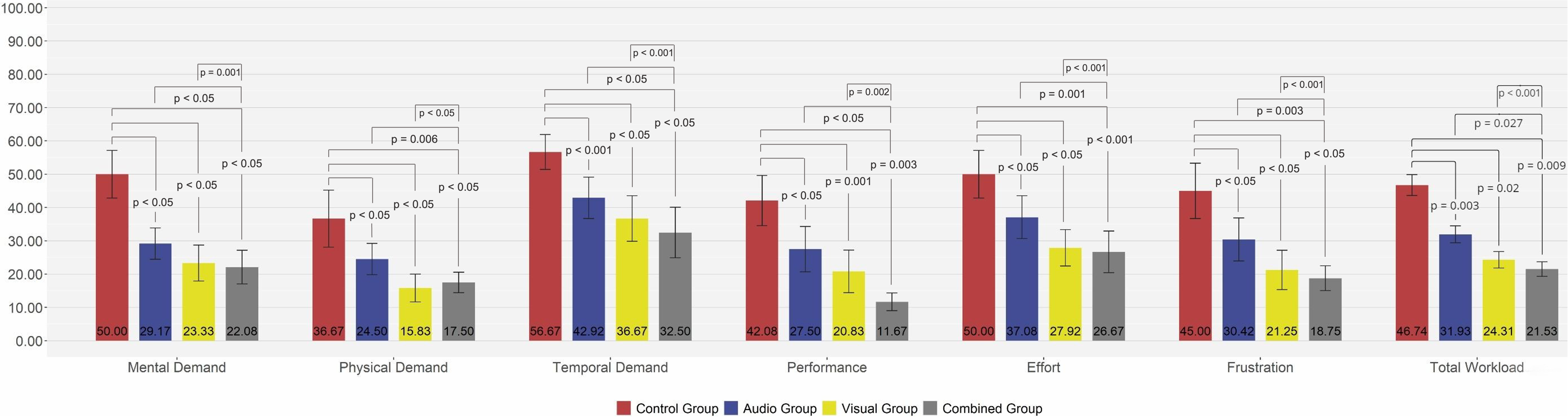}
  \caption{\scriptsize{Second task in Study II: NASA TLX results of control (left) and three experimental groups. Error bars show mean $\pm$ standard error.}}
  \label{fig:II_nasa_task_2}
\end{figure}

The NASA TLX answers gathered from study II are then presented. The descriptive statistics and visualizations of the collective data are shown in Figures \ref{fig:II_nasa_task_1} and \ref{fig:II_nasa_task_2}. By observing the Total Workload, we found that the values of this metric from the control group (Total Workload: $mean$ = 65.33, $SD$ = 9.46 in the first task; $mean$ = 46.83, $SD$ = 7.03 in the second task) were greater than the other three groups for both tasks. Furthermore, participants in the control group yielded the most perceived workload in every index upon every task; in particular, the members of the control group had an evident lead in Mental Demand ($mean$ = 69.17, $SD$ = 17.06), Effort ($mean$ = 77.08, $SD$ = 12.98), and Frustration ($mean$ = 67.50, $SD$ = 16.39) in the first task. For the second task, the differences were not as remarkable; Mental Demand for the control group ($mean$ = 50.00, $SD$ = 24.83) retained the high values. We noticed that for the second task almost all of the workload indexes, including total workload in the combined group were lower than the other three groups; the exception was Physical Demand.

The variance homogeneity was once again revealed in the TLX values. Therefore, the result of the 2*4 mixed ANOVA ($p$ = 0.05) showed that all participants had a statistically significantly lesser perceived workload ($mean$ = 31.25; $SD$ = 11.96) in the second task compared to the first ($mean$ = 46.87; $SD$ = 13.89), $F$(1, 44) = 45.344, $p$ $<$ 0.001. There were significant main effects of the NASA TLX results among the four groups ($F$(3, 44) = 6.049, $p$ = 0.002), but no significant difference was found on the interaction between the tasks and the groups ($F$(3, 44) = 2.387, $p$ = 0.082). For pairwise comparison, the post hoc test showed no significant differences between every pairwise comparison of audio, visual, and combined groups in the first task. However, all the other three groups were significantly different when compared to the control group. That is, the participants from the control group felt that there was significantly more perceived workload than those from the audio group ($mean$ = ; $SD$ = 5.71 $p$ = 0.003), the visual group ($mean$ = 36.50; $SD$ = 10.82; $p$ = 0.001), and the combined group ($mean$ = 45.83; $SD$ = 6.97; $p$ = 0.021). In the second task, the participants from the control group still possessed significantly more perceived workload compared to members of the audio group ($mean$ = 32.00; $SD$ = 6.69; $p$ = 0.003), the visual group ($mean$ = 24.33; $SD$ = 7.31; $p$ = 0.02), and the combined group ($mean$ = 21.83; $SD$ = 7.36; $p$ = 0.009). Nonetheless, the pairs of the audio-combined ($p$ = 0.027) and the visual-combined ($p$ = 0.006) also revealed significant differences. That is, The results from the combined group are statistically different from the other groups in the second task. The result means the synthesis of the two modalities of the hints we tested worked better than when there was only one modal hint in the second task, where the gaze playback was provided as the instant post-task feedback in the same context.

It is noted that the non-control groups had an obvious workload decline compared to the control group in each task whilst the combined group had the lowest perceived workload in the second task (\textbf{H1}) in Study II. All the indexes including Total Workload in every group showed considerable decrease for the second task (\textbf{H2}).

\subsection{Summary of Qualitative Analysis}

\subsubsection{Hints}
Most participants expressed compliments for both the visual and audio hints in the process of searching the books with the AR glasses mounted. The participants mostly stated that they obtained much help from the visual hint in guiding them during the searching procedure. One participant in the visual group from Study I said: \textit{“The arrow was brightly, clearly, and precisely pointing out the correct shelf which guided me finding the book.”} Another participant in the visual group from Study II explained: \textit{“It was great to see the visual arrow appear immediately because I was expecting something slow. Also, the arrow sign was big so that I could find the book in a time-saving way.”} One participant from the combined group in Study II described the high comprehensibility of the hint: \textit{“It was easy to understand the meaning of the arrow that facilitated the pace of my book searching.”} The audio hint was praised mainly due to its unambiguity and clearness. One participant from the audio group in Study I reported: \textit{"the auditory message was an excellent directive in leading me to the target shelf. It was a very clear information with an appropriate rate that I could catch the main content smoothly."} Another participant from the combined group in Study II said: \textit{"The hint was clear and concise which was easy to follow. It took little time to comprehend so that it was helpful for me to set the goal during searching the book."}

There were also some critiques for the two types of hints. One participant from the visual group in Study I said: \textit{“It was a bit difficult to notice there was a visual arrow since it was out of my sight, but it turned out to be working after I realized that. The arrow could be put on the position closer to the centre of the sight.”} For the audio hint, one participant from the combined group in Study II suggested \textit{“I would have missed the audio instruction since it was a short message, which needed me to concentrate more on the task. It would be better if the hint was followed by one reminder on the screen, such as "please be aware of the incoming audio hint".”}

\subsubsection{Task Feedback}
Only the 48 participants in Study II experienced the instant post-task feedback. Most of them gave a positive appraisal of the gaze playback. For example, one participant commented: \textit{"The gaze playback was a good feedback since it displayed to me how I was searching for the book in the first task, which made me get better self-organized on how I tried to locate the book in the second task."} Another participant told us: \textit{"the gaze playback was amazing since it told me do not look at other irrelevant places when searching the second book."}. However, one participant also criticized: \textit{"the dot fluctuated so quickly that I could not follow and understand what it intended to tell me. It could be better if the gaze trajectory line was also visualized."}

Overall, both the visual/audio hints and instant post-task feedback were regarded helpful by all participants. When checking the feedback given by them, we realized that the visual arrow and audio message were playing a dominant role in guiding the participants to the targets, which substantially shortened the task periods and decreased the workload. We also believe that after reviewing the gaze playback, the participants gained more awareness of how to perform and improve the book-searching task.

\subsection{Findings}
By analyzing all the results from the two studies, we found that the TCT and task perceived workload reported gave identical results thus confirming our hypotheses. The visual/audio/combined groups all had a significant difference from the control group in both TCT and NASA TLX, while the two tasks were significantly different when coupling with the instant post-task feedback. In addition, the combined group had significant differences with both visual and audio groups in Study II where the feedback was supplied. We believe the results obtained are representative and can be harnessed to verify our hypotheses. Thus, we rephrase our findings as follows since \textbf{H1} was partially verified while \textbf{H2} and was completely affirmed.

\begin{itemize}
    \item Even though visual and audio hints both have a positive effect in facilitating task performance and reducing perceived workload in AR visual search tasks, the combination of these two hints has a greater effect than either does individually \textbf{under the condition that there is instant post-task feedback}.
    \item Instant post-task feedback through gaze assistance has the capacity to stimulate better performance and reduce perceived workload in the same context.
\end{itemize}

\section{Discussion and Limitations}
\label{dis}

\subsection{Insights}
What did we do and achieve in our experiments? We started from two core hypotheses, \textbf{H1} and \textbf{H2}, with the purpose of discovering the effects of visual/audio hints in either single or combined setups for AR search tasks. We also studied the influence of instant post-task feedback in the same context. To resolve the assumptions, we developed an AR approach engaging two modules of the hints and gaze-assisted instant post-task feedback separately. We designed a case study for a visual book-searching task which involved the use of a mobile AR headset. An AR app was installed in Microsoft HoloLens 2. We first ran a pre-testing session to assure the functionality of our AR solution, as well as to gain early-stage feedback. After some adjustments, we designed and implemented a more comprehensive, larger-scale user study, where 96 participants were allocated in Study I (n=48) and II (n=48) and placed randomly four groups (control (n=12), audio (n=12), visual (n=12), and combined (n=12)) in each sub-study. The user study was based on the between-subject principle with within-subject factor and the collected results were analyzed correspondingly. Participants were extensively engaged in searching tasks within two distinct bookcases, each containing a diverse range of totally different books, ensuring the adequacy of the study for deriving meaningful results. Moreover, we have taken into account the exposure time of participants to experimental conditions, assuring that it does not compromise the validity of the results. Throughout the entire study and data analysis, \textbf{H1} was partially verified, while \textbf{H2} was completely verified. 

Our proposed solution proves to be advantageous for increasing AR visual search task performance and simultaneously reducing the perceived workload on users. We found that the combination of visual and audio hints can ameliorate both task performance and perceived workload when coupled with instant post-task feedback. We think that participants were mentally self-reflected and guided by the feedback regarding their searching modes during the tasks, as mentioned in qualitative results. A few reported that they improved the task performance by slightly changing the searching mode/route in the second task. Also, the "double guidance" brought by the combination of visual and audio hints positively affected participants after their self-reflection from the post-task feedback. We believe our results are promising, both encouraging further investigations in the field of using directive hints within mobile HMD setups, and incorporating an immediate opportunity for user self reflection and evaluation. The noticeable decrease of TCT and the cognitive task workload was a confirmation of our hypotheses. Beyond the fundamental results obtained, the research implication of our study has high generalizability and convertibility. We aimed to benefit the general AR searching specialization by selecting the visual book-searching task since all the essential elements needed for visual searching were contained in this use case.

\subsection{Limitations}
We acknowledge that our study has some limitations. First, we designed two visual book-searching tasks to be performed using two bookcases that contained different books, eliminating memory as a factor in the second task. 
Second, The books themselves can serve as the variable elements. The thickness and order of the books might alter the participants' search speeds and workload. The books we chose were generally thick ones, and the participants' searches might have been sped up to some extent because the title displayed on the spine would be larger. Also, the books' positioning may affect participants' searching times since individuals can have varied viewing habits. A book put in the middle, for example, might be more easily noticed by people who are used to searching from the middle; a person who was used to searching from left to right might find a book on the left faster in the first task, while taking longer to find a book located on the right in the second task. Additionally, some books having similar titles might disturb and confuse participants which may lead to a longer or incorrect searching process. Nonetheless, there could potentially exist ambient visual or auditory disruptions (such as the objects except books on the bookcases and unforeseen background noises) that might have a slight impact on the participants.

Third, the inclusion of participants of our study might not be sufficient. A more diverse group of samples could be considered in the future to strengthen the reliability of the research conclusions. For example, we did not yet consider incorporating non-binary or LGBTQ+ participants \cite{walker2020more}. Fourth, we employed solely the NASA TLX as the subjective evaluation measure. Other subjective questionnaires could be utilized to incorporate more diverse subjective aspects, such as satisfaction and pleasure. Fifth, due to the performance sequence, the gaze review only influenced the second task in our study. However, in order to better understand the influence of gaze review on search tasks, there could be another design solution with a counterbalancing factor \cite{bradley1958complete} where the gaze playback could be recorded from the second task then used for the first task.

Finally, there is room for improvement in the AR app, especially for the design of the visual.audio hints. The arrow used in the current app was calibrated to the position of the HMD. Thus, the relative heights of arrows and the bookcases differed due to the different heights of the participants. As a result, an arrow pointing to a shelf might go slightly up or down. Participants in the visual group might perceive higher workload and spend more time aligning the arrow with a specific shelf due to inaccurate calibration. To address this, future work should calibrate the arrow positions with the bookcases rather than the HMD itself. In addition, the visual arrow used in our study led to advisable results but it might be too monotonous. Different design principles of the visual hints should be considered, e.g., not only the arrows but also other types of graphic representations which can visually guide participants. Furthermore, more options of the color rendering, the size, and the positioning of the visual hints should be discussed and compared to make the proof-of-concept more robust.

\section{Conclusions and Future Work}
\label{con}
In this paper, we have described an AR approach for visual search tasks with supported visual/audio hints and gaze-assisted instant post-task feedback based on mobile HMD. Previous research had given us an incentive to investigate the effect of hints and feedback. Based on our hypotheses, we designed and conducted a case study of visual book-searching where the gaze playback acted as the instant post-task feedback based. The experimental procedure consisted of a comprehensive user study (n=96) with two comparative sub-studies. The resulting analysis was founded mainly on collected NASA TLX answers with TCT measurement as a preliminary analytic metric. We partially verified \textbf{H1}: both visual and audio hints have a positive effect in facilitating task performance. The combination of the two hints works better than either individually, under the condition that there is instant post-task feedback. \textbf{H2} was completely verified: instant post-task feedback has the capacity to bring about better performance. We pointed out the high generalizability and convertibility of our research output in making advantage of the general AR searching processes.

Our research is novel because it fills the gaps of combining visual and audio hints in the same AR context, the integration of feedback and gaze assistance. More importantly, it not only focuses on the AR hints and instant post-task feedback, but places them in the same context and generates meaningful conclusions for universal AR visual search tasks. In future work, we will make our app more adjustable for human eyes and thereby user-friendly by shortening users' adaptation time. We will also improve our system by adding adaptive support which can react to users' gaze by giving additional help to users focusing too long on the wrong parts of the environment. Next, we intend to involve more types of visual/audio hints, as well as more modalities of feedback. Meanwhile, more demographic information will be collected and more effort will be made to involve LGBTQ+ participants. We foresee carrying out a more inclusive user study with more metrics for evaluation. We hope that our research will have an impact on industry-based AR setups and configurations empowering people's capabilities and levels of efficiency in general AR search tasks.

\acknowledgments{%
This work was partly supported by the Norges Forskningsrad (309339, 314578), MediaFutures user partners and Universitetet i Bergen.
}

\bibliographystyle{abbrv-doi-hyperref}

\bibliography{template}

\end{document}